\documentclass{article} % [a4paper][11pt]  

\usepackage{url}
\usepackage{alltt}
\usepackage{amsmath} 
\usepackage{amssymb}

\usepackage{epsfig}
\usepackage{wrapfig}
\usepackage{subfigure}

\usepackage{rotating}
\usepackage{moreverb}
\usepackage{fancyvrb}

\newcommand{\qed}{\hfill \ensuremath{\Box}}
\newtheorem{definition}{Definition}
\newtheorem{theorem}{Theorem}
\newtheorem{lemma}{Lemma}
\newtheorem{example}{Example}

\def\bd{\begin{description}}
\def\ed{\end{description}}
\def\bc{\begin{center}}
\def\ec{\end{center}}
\def\bq{\begin{quote}}
\def\eq{\end{quote}}
\def\bi{\begin{itemize}}
\def\ei{\end{itemize}}
\def\be{\begin{enumerate}}
\def\ee{\end{enumerate}}
\def\ba{\begin{array}}
\def\ea{\end{array}}

\newcommand{\chr}{{CHR}}
\newcommand{\true}{{\it true}}

\newcommand{\false}{{\it false}}

\newcommand{\CT}{\ensuremath{\mathcal{CT}}}

\newcommand{\simp}{{\; \Leftrightarrow \;}}

\newcommand{\myparagraph}[1]{\textbf{#1.}}

\begin{document}

\title{Repeated Recursion Unfolding for Super-Linear Speedup within Bounds}  %!!!
\author{%
Thom Fr{\"u}hwirth\\
Ulm University, Germany\\
\tt{thom.fruehwirth@uni-ulm.de}
}

\maketitle

{\sf \noindent This is the full version of a paper presented at 
the 30th International Symposium on Logic-Based Program Synthesis and Transformation (LOPSTR 2020).}

\begin{abstract}

Repeated recursion unfolding is a new approach that repeatedly
unfolds a recursion with itself and simplifies it while keeping all unfolded rules. 
Each unfolding doubles the number of recursive steps covered.
This reduces the number of recursive rule applications to its logarithm at the expense of introducing a logarithmic number of unfolded rules to the program.

Efficiency crucially depends on the amount of simplification inside the unfolded rules. 
We prove a super-linear speedup theorem in the best case, 
i.e. speedup by more than a constant factor.
Our optimization can lower the time complexity class of a program.
In this paper, the super-linear speedup is within bounds: it holds up to an arbitrary but chosen upper bound on the number of recursive steps.

We also report on the first results with a prototype implementation of repeated recursion unfolding.
A simple program transformation completely removes recursion up to the chosen bound. The actual runtime improvement quickly reaches several orders of magnitude.

\medskip
\myparagraph{Keywords}
Program Transformation, Program Optimization, Super-Linear Speedup, Recursion, Speedup Theorem, Time Complexity.
\end{abstract}

\section{Introduction}\label{intro}

In the context of rule-based and logic programming,
unfolding is the program transformation that replaces a call (goal) in the body of a rule by the body of a rule whose head is matched by the goal. 
This saves a rule application at runtime. More importantly, the resulting code can be simplified for efficiency.

\begin{example}[Summation]
Consider the following simple recursive program
written in CHR (Constraint Handling Rules). It adds all numbers from $1$ to $n$.
\begin{gather*}
r_0: sum(N,S) \simp N>1 \, |\, S := N{+}S1, sum(N{-}1,S1)\\
   sum(N,S) \simp N=1 \, |\, S=1.
\end{gather*}
Head, guard and body of a rule are separated by $\simp$ and $|$, respectively.
The rules can be understood as a procedure definition for $sum$, each rule covering a case.
When a call matches the head of a rule and the guard condition holds, the body of the rule is executed.
In CHR code, upper case letters stand for logical variables.

Unfolding the recursive rule with itself will result in
\begin{gather*}
r_1: sum(N,S) \simp N>1, N{-}1>1 \, |\, S:=N{+}S1, N'=N{-}1, 
S1:=N'{+}S1',\\
 sum(N'{-}1,S1'),
\end{gather*}
which can be simplified to
$$r_1: sum(N,S) \simp N>2 \, |\, S := 2{*}N{-}1{+}S1', sum(N{-}2,S1').$$
Note that this rule $r_1$ cannot replace the original recursive rule because it only applies in case $N>2$. It behaves like applying the original rule $r_0$ twice. 
It saves computation time, because 
we only need about half as many recursive steps as with the original rule alone. 
Since the simplification of the unfolded rule results in code of similar size and time complexity as the original rule, we can expect to halve the runtime.

If we would now unfold the recursive goal in rule $r_1$ with rule $r_0$ we have unfolded thrice overall. We can expect a speedup of roughly three times if we replace $r_1$ by the new rule. But we can do even better than that
if we keep rule $r_1$ and unfold rule $r_1$ with itself. 
The resulting rule after simplification is
$$r_2: sum(N,S) \simp N>4 \, |\, S := 4*N{-}6+S1, sum(N{-}4,S1).$$
This rule results in a four-fold speedup.
\end{example}

We can repeat this process: unfolding the newly generated recursive rule with itself until we reach a given upper bound. 
We never unfold with the base case. It is ignored.
This is what we call {\em repeated recursion unfolding}.
With each unfolding, the number of original recursive steps simplified into one recursive step of the unfolded rule will double.

Given $n$ recursive steps with the original rule and $log_2(n)$ unfolded rules according to our scheme, it is possible to have at most $log_2(n)$ recursive rule applications instead of $n$.
If the unfolded rules take not more time than the original recursive rule,
the time complexity class can be lowered and super-linear speedup in runtime can be achieved.

For our scheme to work, we always have to apply the best, most unfolded rule. To implement this behavior, we use another program transformation that removes the recursion altogether. We call this {\em recursionless recursion}.

\myparagraph{Overview of the Paper}
Section \ref{prelim} introduces syntax and semantics of the CHR programming language using a single state transition.

Section \ref{main} defines our program transformation scheme of repeated recursion unfolding with simplification and proves its correctness.
We will use summation as our running example.

Section \ref{sectime} proves that there exists a simple optimal rule application strategy for unfolded recursions with best-case simplification that results in super-linear speedup by lowering the time complexity classes by a factor of up to $O(n)$.

Section \ref{secrecno} introduces recursionless recursion, a transformation that provides a semi-naive implementation of optimal rule applications. We prove that the worst-case overhead of this scheme
is linear in the number of unfolded rules.

In Section \ref{secbench} we conclude our summation example with benchmarks in our prototype implementation. It improves time complexity from linear to constant.
Another example, naive list reversal, improves from quadratic to linear complexity. It runs faster than the hand-optimized built-in list reversal in Prolog, the implementation language of CHR.

Finally, we discuss related work and end with conclusions.

\section{Preliminaries}\label{prelim}

We recall the abstract syntax and the equivalence-based abstract operational semantics of CHR (Constraint Handling Rules) \cite{fru_chr_book_2009} in this section. We also informally describe the refined operational semantics typically realized in sequential implementations of CHR.

\subsection{Abstract Syntax of CHR}

The CHR language is based on the abstract concept of constraints.
{\em Constraints} are relations, distinguished predicates of first-order predicate logic.
There are two kinds of constraints: {\em built-in (pre-defined) constraints} 
and
{\em user-defined (CHR) constraints} 
which are defined by the rules in a CHR program.
Built-in constraints can be used as tests in the guard as well as for auxiliary computations in the body of a rule.
There are at least the built-in 
constraints $\true$ and $\false$, syntactical equality $=$ over finite terms
and the usual relations over arithmetic expressions.
Upper-case letters stand for (possibly empty) conjunctions of constraints in definitions, lemmas and theorems.
\begin{definition}[CHR Program and Rules]
{%\rm
A {\em \chr\ program} is a finite set of rules.  
A {\em (generalized) simplification rule} is of the form
\[r: H \Leftrightarrow C \, |\, B,\]
where $r$ is an optional {\em name} (a unique identifier) of a rule.
The {\em head} $H$ is a conjunction of user-defined constraints,
the optional {\em guard} $C$ is a conjunction of built-in constraints,
and the {\em body} $B$ is a goal.

The {\em local variables of a rule} are those not occurring in the head of the rule.
A {\em renaming (variant, copy)} of a goal is obtained by uniformly replacing its variables by other variables.

Conjunctions are understood as multisets of their atomic conjuncts.
We often use simple commas to denote logical conjunction to avoid clutter.
A {\em goal} is a conjunction of built-in and user-defined constraints.
A {\em state} is a goal.
} %rm
\end{definition}

\subsection{Abstract Operational Semantics of CHR}\label{sec:chr:semantics}

Computations in CHR are sequences of rule applications. The operational semantics of CHR is given by the state transition system. 
It relies on a structural equivalence between states that abstracts away from technical details in a transition \cite{raiser_betz_fru_equivalence_revisited_chr09,betz2014unified}.

{\em State equivalence} treats built-in constraints semantically and user-defined constraints syntactically.
Basically, two states are equivalent if their built-in constraints are logically equivalent (imply each other) and 
their user-defined constraints form syntactically equivalent multisets in this context.
For example, 
$$X{=<}Y \land Y{=<}X \land c(X,Y) \ \equiv \ X{=}Y \land c(X,X) \not\equiv X{=}Y \land c(X,X) \land c(X,X).$$

Let $\CT$ be a (decidable) constraint theory for the built-in constraints.  
\begin{definition}[State Equivalence]
{%\rm
\cite{raiser_betz_fru_equivalence_revisited_chr09}
Let $C_i$ be the built-in constraints, let $B_i$ denote user-defined constraints, and let ${\cal V}$ be a set of variables.
Variables of a goal or state that do not occur in ${\cal V}$ are called {\em local variables of the goal or state}.
Two states $S_1 = (C_1 \land B_1)$ and $S_2 = (C_2 \land B_2)$
with local variables $\bar x$ and $\bar y$ that have been renamed apart
are {\em equivalent}, 
written $S_1 \equiv_{\cal V} S_2$, if and only if
$$	\CT 
\models 
        \forall (C_1 \rightarrow \exists \bar y ((B_1 = B_2) \land C_2))
	\land %\\
	%\indent \indent 
         \forall (C_2 \rightarrow \exists \bar x ((B_1 = B_2) \land C_1))
$$
} %rm
\end{definition}
Note that this definitions implies 
$$\CT \models 
\forall (\exists \bar x (B_1 \land C_1) \leftrightarrow \exists \bar y (B_2 \land C_2).$$
It also makes sure that there is a one-to-one correspondence between user-defined constraints as enforced by $B_1 = B_2$. 
It allows the renaming of local variables.
Local variables can be removed if logical equivalence is maintained.
Occurrences of local variables can be substituted by terms if logical equivalence is maintained.
These properties have been shown in \cite{raiser_betz_fru_equivalence_revisited_chr09}.
An example illustrates these properties of state equivalence and 
the effect of the global variables ${\cal V}$:
$$X{=}Y \land c(X,Y) \equiv_{\{X\}} c(X,X) 
\mbox{ but } 
X{=}Y \land c(X,Y) \not\equiv_{\{X,Y\}} c(X,X).$$

We may drop ${\cal V}$ from the equivalence if it is clear from the context.

Using this state equivalence, the abstract CHR semantics is defined by a single transition
(computation step) between states. It defines the application of a rule. 
If the source state can be made equivalent to a state that contains the head and the guard of a renaming of a rule, then we can apply the rule by replacing the head by the body in the state. Any state that is equivalent to this target state is also in the transition relation.
\begin{definition}[Transition]\label{transition}
{%\rm
A CHR {\em transition (computation step)}
$S \mapsto_r T$ is defined as follows, where $S$ is called {\em source state} and $T$ is called {\em target state}:

\begin{center}
$\underline{S \equiv_{\cal V} (H \land C \land G)  \not\equiv \false \  \ (r : H \Leftrightarrow C \, |\, B) \  \ (C \land B \land G) \equiv_{\cal V} T}$\\
$S \mapsto_r T$
\end{center}
where the rule $(r : H \Leftrightarrow C \, |\, B)$ is a renaming
of a rule from a given program $\cal{P}$
such that its local variables do not occur in $G$. %!!!!

A {\em computation (derivation)} of a goal (query) $S$ with variables ${\cal V}$ in a program $\mathcal{P}$
is a connected sequence
$S_i \mapsto_{r_i} S_{i+1}$ beginning with
the {\em initial state (query)} $S_0$ that is $S$
and ending in a {\em final state (answer, result)} $S_n$ or 
otherwise the sequence is infinite and the computation is {\em non-terminating (diverging)}.
We may drop the reference to %the program $\mathcal{P}$ and 
the rules.
The relation $\mapsto^*$ denotes the reflexive and transitive closure of $\mapsto$.
} %rm
\end{definition}
The goal $G$ is called {\em context} of the rule application. It remains unchanged. 
It may be empty.
Note that CHR is a committed-choice language, i.e. there is no backtracking or undoing of rule applications.

\subsection{Refined Operational Semantics of \chr}\label{sec:chr:refsemantics}

Almost all sequential CHR
implementations execute queries and rule body constraints from left to right and apply rules top-down following
their textual order of the program. This behavior has
been formalized in the so-called refined semantics which is a
concretization of the abstract operational semantics \cite{duck_stuck_garc_holz_refined_op_sem_iclp04}.

In this refined semantics, a CHR constraint in a goal
can be understood as a procedure call that goes efficiently through the rules of the program.
If the current goal matches the head constraints of a rule
and if, under this
matching, the guard check of the rule holds in the current context, the rule is applicable.
Given a query, the rules of the program are applied to exhaustion. 
When a simplification rule is applied, the matched constraints are replaced by the body of the rule.

\section{Repeated Recursion Unfolding}\label{main}

We recall a definition of rule unfolding in CHR, then define and prove correctness of simplification inside rules in order to introduce repeated recursion unfolding.

\subsection{Rule Unfolding}

For correct unfolding of rules in CHR, we follow the definition of \cite{gabbrielli2015unfolding},
where also correctness of the unfolding function is shown. 
This means that we can safely add the unfolded rule to a program while preserving its semantics. In other words, a correctly unfolded rule is always redundant (but, of course, is expected to improve efficiency).
In this paper we specialize the definition to the case of CHR simplification rules instead of arbitrary CHR rules. This simplifies the definition and is sufficient for our purposes.

For a goal $A$,
let $vars(A)$ denote the set of variables in $A$.
A substitution is based on a mapping function from variables to terms 
$\theta: {\cal V} \rightarrow {\cal T}$, written in postfix notation,
such that domain of $\theta$,
the set $dom(\theta) = \{X \mid X\theta \neq X\}$ is finite.
When a substitution is applied to a goal, it is applied to all variables in the goal.
If $A=B\theta$, where $B$ is a goal, we say that $A$ is an {\em instance} of $B$,
$A$ {\em matches} $B$, and that $B$ is {\em instantiated}.

Set difference $C_1 = C_2 \setminus C_3$ for built-in constraints is defined
as $C_1 = \{c \in C_2 \mid \mathcal{CT} \not \models C_3 \rightarrow c\}$.
In words, $C_1$ does not contain the constraints from $C_2$ that are implied by $C_3$.

\begin{definition}[Unfolding]\label{def:unf}
(Def. 8 \cite{gabbrielli2015unfolding})
{%\rm
Let ${\cal P}$ be a CHR program and let $r, v \in {\cal P}$ be two
rules whose variables have been renamed apart
$$
\begin{array}{lcl}
r: H & \Leftrightarrow &  C\,|\, D \land B \land S\\
v: H' & \Leftrightarrow & C '\,|\, B',
\end{array}
$$
where $D$ is the conjunction of the built-in constraints in the body of $r$. 
Then we define 
$$\mathit{unfold}(r,v) = r'$$ 
as follows.
Let  $\theta$ be a substitution such that 
$dom(\theta) \subseteq vars(H')$ and
$\mathcal{CT}  \models (C \wedge D) \rightarrow S{=}H'\theta$,
then the unfolded rule $r'$ is:
$$r': H \Leftrightarrow C\land C''\theta \, |\, D \land B \land S{=}H' \land B',$$ 
where 
$C''\theta = C'\theta \setminus (C \wedge D)$ with
$vars(C''\theta) \cap vars(H'\theta) \subseteq vars(H)$ 
and $\mathcal{CT} \models \exists (C \wedge C''\theta)$.
} %rm
\end{definition}
If a goal $S$ in the body of rule $r$ matches the head $H'$ of a rule $v$,
unfolding replaces $S$ in the body of rule $r$ by the body of rule $v$ together with $S{=}H'$ to obtain a new rule $r'$. In the resulting rule $r'$ we also add to its guard $C$ an instance of a part of the guard of rule $v$. 
This part $C''$ contains the non-redundant built-in constraints that are not already implied by the built-in constraints in the guard and body of the rule $r$. 

Note that for a correct unfolding according to the above definition,
three conditions have to be satisfied. 
First,
$$\mathcal{CT}  \models (C \wedge D) \rightarrow S{=}H'\theta,$$
means that goal $S$ must match the head $H'$, i.e. be an instance of $H'$, in the context of the built-in constraints of the rule $r$. This condition reflects the fact that the rule $v$ should be applicable to $S$.

Second, the non-obvious condition
$$vars(C''\theta) \cap vars(H'\theta) \subseteq vars(H)$$
means that for correctness the variables shared between the instantiated head $H'\theta$ and instantiated simplified guard $C''\theta$ of the rule $v$ must also occur in $H$.
Note that using $C''$ instead of $C$ can make the set of variables smaller that have to occur in $H$.

Third, the satisfiability of the guard of the unfolded rule $r'$, $\mathcal{CT} \models \exists (C \wedge C''\theta)$, ensures that the rule is nontrivial in that it has a satisfiable guard. Otherwise the rule would never be applicable.

\begin{example}[Summation, contd.]{%\rm
We unfold the recursive rule for summation with (a copy of) itself:
\begin{gather*}
r: sum(N,S) \simp N>1 \, |\, S := N{+}S1, sum(N{-}1,S1)\\
v: sum(N',S') \simp N'>1 \, |\, S' := N'{+}S1', sum(N'{-}1,S1').
\end{gather*}
Then the unfolded rule is
\begin{gather*}
sum(N,S) \simp N>1, N{-}1>1 \, |\, S:=N{+}S1, sum(N{-}1,S1){=}sum(N',S'),\\ 
S':=N'{+}S1', sum(N'{-}1,S1').
\end{gather*}
Unfolding is correct since its three conditions are satisfied.
First, $sum(N{-}1,S1)$ is an instance of $sum(N',S')$, i.e.
$$(N>1, S := N+S1) \rightarrow sum(N{-}1,S1)=sum(N',S')\theta,$$
since $\theta$ can map $N'$ to $N{-}1$ and $S'$ to $S1$.
Second, 
$$vars(N{-}1>1) \cap  vars(sum(N{-}1,S1)) \subseteq vars(sum(N,S))$$ 
holds since $\{N\}  \cap  \{N,S1\}\subseteq \{N,S\}.$
Third, the new guard $N>1, N{-}1>1$ is satisfiable.
} %rm
\end{example}
Obviously we can simplify the built-in constraints of the guard and the body of this rule, and we will define this kind of simplification next.

\subsection{Rule Simplification}\label{rulesimp}

Speedup crucially depends on the amount of constraint simplification that is possible in the unfolded rules. 
The goal is to replace constraints by semantically equivalent ones that can be executed more efficiently.

\begin{definition}[Simplification]\label{def:simp}
{%\rm
Given a rule $r$ of the form
$$r: H \Leftrightarrow C \, |\, D \land B,$$
where $D$ are the built-in constraints and $B$ are the user-defined constraints in the body of the rule.
We define
\begin{gather*}
\mathit{simplify}(r) = 
(H' \Leftrightarrow C' \, |\, D' {\setminus} C' \land B') \mbox{ such that}\\
(H \land C) \equiv_{{\cal V}} (H' \land C')
 \mbox{ and } (C \land D \land B) \equiv_{{\cal V}} (D' \land B'),
\end{gather*}
where $D'$ are the built-in constraints and $B'$ are the user-defined constraints in the body of the rule
and
where ${\cal V} = vars(H) \cup vars(H')$.
} %rm
\end{definition}

In the given rule, we replace head and guard, and the body, respectively, by simpler yet state equivalent goals.
The choice of ${\cal V}$ allows us to remove local variables if possible, i.e those that occur only in the guard or body of the rule.
We temporarily add the guard $C$ when we simplify the body to ensure correctness and improve the simplification.

For correctness we have to show that the same transitions $S \mapsto T$ are possible with rule $r$ and rule $\mathit{simplify}(r)$.

\begin{theorem}[Correctness of Rule Simplification]
{%\rm
Let $r = (H \Leftrightarrow C \, |\, D \land B)$ be a rule and
let $s = (H' \Leftrightarrow C' \, |\, D' {\setminus} C' \land B')$ be the simplified rule $\mathit{simplify}(r)$.
For any state $S$ and variables ${\cal V}$, $S \mapsto_r T$ iff $S \mapsto_s T$.

\myparagraph{Proof}
According to the definition of a CHR transition and $\mathit{simplify}(r)$, we know that
\begin{center}
$S \mapsto_r T \mbox{ iff }
S \equiv_{\cal V} (H \land C \land G)  \not\equiv \false \mbox{ and } (C \land D \land B \land G) \equiv_{\cal V} T$\\
$S \mapsto_s T \mbox{ iff }
S \equiv_{\cal V} (H' \land C' \land G')  \not\equiv \false \mbox{ and } (C' \land D' {\setminus} C' \land B' \land G') \equiv_{\cal V} T$\\
$(H \land C) \equiv_{\cal V'} (H' \land C')$\\
$(C \land D \land B) \equiv_{\cal V'} (D' \land B').$
\end{center}
It suffices to show that $S \mapsto_r T$ implies $S \mapsto_s T$, since the implication in the other direction is symmetric and can be shown in the same way.
Hence we have to show that there exists a goal $G'$ such that
\begin{center}
$S \equiv (H \land C \land G) \equiv_{\cal V} (H' \land C' \land G') \mbox{ if }
(H \land C) \equiv_{\cal V'} (H' \land C')$ and\\
$T \equiv (C \land D \land B \land G) \equiv_{\cal V} (C' \land D' {\setminus} C' \land B' \land G')  \mbox{ if } (C \land D \land B) \equiv_{\cal V'} (D' \land B').$
\end{center}
We choose $G' = C \land G$. %!!!!
The full proof can be found in the appendix. %full version of the paper that is online.
\qed
}%rm
\end {theorem}

\begin{example}[Summation, contd.]{%\rm
Recall the unfolded rule
\begin{gather*}
sum(N,S) \simp N>1, N{-}1>1 \, |\, S:=N{+}S1, sum(N{-}1,S1){=}sum(N',S'),\\ 
S':=N'{+}S1', sum(N'{-}1,S1').
\end{gather*}
For the head and guard we have that
$$sum(N,S), N>1, N{-}1>1 \equiv_{\{S,N\}} sum(N,S), N>2.$$
For the body we have that
\begin{gather*}
N>1, N{-}1>1, S := N{+}S1, sum(N{-}1,S1){=}sum(N',S'), S' := N'{+}S1',\\sum(N'{-}1,S1') 
\equiv_{\{S,N\}} 
N>2, S := 2{*}N{-}1{+}S1', sum(N{-}2,S1').
\end{gather*}
Thus the rule can be simplified into the rule
$$sum(N,S) \simp N>2 \, |\, S := 2{*}N{-}1{+}S1', sum(N{-}2,S1').$$
} %rm
\end{example}

\subsection{Repeated Recursion Unfolding}

We now define what it means to unfold a given recursive rule with itself, to simplify it, to repeat this process, and to add the resulting rules up to a given bound to the original program.

\begin{definition}[Repeated Recursion Unfolding]\label{def:rru}
{%\rm
Let $r$ be a recursive rule in a given program ${\cal P}$. 
The {\em unfolding of a recursive rule} $r$ is defined as
$$\mathit{unfold}(r) = \mathit{unfold}(r,r)$$

The {\em repeated unfolding} of a recursive rule $r$ with simplification
is a sequence of rules $r_0, r_1, \ldots, r_i, \ldots$ where
\begin{gather*}
r_0 = r \in {\cal P}\\
r_{i+1} = \mathit{simplify}(\mathit{unfold}(r_i))
\end{gather*}

Let $n$ be an upper bound on the number of recursive steps (recursion depth)
for rule $r$. The {\em recursively unfolded program}
${\cal P}^{r,n}$ of rule $r$ is defined as
$${\cal P}^{r,n} = {\cal P} \cup \bigcup_{i=1}^{\lfloor\log_2 (n)\rfloor} r_{i}$$ 

} %rm
\end{definition}

\begin{example}[Summation, contd.]{%\rm
Recall the unfolded simplified rule
$$sum(N,S) \simp N>2 \, |\, S := 2{*}N{-}1{+}S1, sum(N{-}2,S1).$$
We repeat the unfolding:
\begin{gather*}
sum(N,S) \simp N>2, N{-}2>2 \, |\, S := 2{*}N{-}1{+}S1,\\
sum(N{-}2,S1){=}sum(N',S'), 
S' := 2{*}N'{-}1{+}S1', sum(N'{-}2,S1').
\end{gather*}
The unfolded rule can be simplified into the rule
$$sum(N,S) \simp N>4 \, |\, S := 4{*}N{-}6{+}S1', sum(N{-}4,S1').$$
The complete program for repeated recursion unfolding $sum$ three times is:
\begin{gather*}
sum(N,S) \simp N>8 \, |\, S := 8*N{-}28+S1, sum(N{-}8,S1)\\
sum(N,S) \simp N>4 \, |\, S := 4*N{-}6+S1, sum(N{-}4,S1)\\
sum(N,S) \simp N>2 \, |\, S := 2*N{-}1+S1, sum(N{-}2,S1)\\
sum(N,S) \simp N>1 \, |\, S := N+S1, sum(N{-}1,S1)\\
sum(N,S) \simp N=1 \, |\, S=1.
\end{gather*}
Note that for the query $sum(9,R)$ we could use any of the recursive rules with the same result. Of course the most efficient way is to use the first rule.
We will discuss optimal rule applications in the next section.
} %rm
\end{example}

\section{Time Complexity and Super-Linear Speedup}\label{sectime}

We first show that we can save on rule applications with our recursive unfolding scheme:
we will always apply the most unfolded recursive rule to perform a maximum number of recursive steps with a minimum number of rule applications.
Significant speedup then crucially depends on the amount of constraint simplification that is possible in the unfolded rules. 
Then we show that simplification leads to super-linear speedup in the best case, i.e. a change into a lower time complexity class.

\subsection{Optimal Rule Applications}

By definition of correct unfolding, the unfolded rule is redundant with regard to the original rule. The other direction is not necessarily true. 
Even if the number of recursive steps should admit the application of an unfolded rule instead of the original rule, it may not be possible because the guard of the unfolded rule may be somewhat too strict.

\begin{lemma}[Optimal Rule Applications]\label{optrule}
{%\rm
Given a program ${\cal P}$ with a recursive rule $r$.
Assume that the unfolded rules in the unfolded program ${\cal P}^{r,n}$
satisfy the following condition:
If a rule $r_i$ with $i < log_2(n)$
can perform two recursive computation steps in a state, then 
rule $r_{i+1}$ can perform one computation step.

Given a computation with the rule $r$ that takes $n$ recursion steps.
Then there exists a computation with at most $log_2(n)$ recursive rule applications in the unfolded program ${\cal P}^{r,n}$
that results in an equivalent state. We call these {\em optimal rule applications}.

\myparagraph{Proof}
The original recursive rule $r$ performs one recursive step when applied. 
With each unfolding, the number of steps covered by the new unfolded rule is doubled. 
For rule $r_i$, $2^i$ recursive steps are covered.

From the condition in the claim it follows that for a query with recursion depth $n$, any unfolded rule $r_i$ is applicable
with $n \geq 2^i$. 
Since the unfolding is correct, the resulting states will be equivalent.
In one application step of rule $r_i$, the recursion depth will be reduced to $n-2^i$.  

Then the {\em optimal rule application strategy} is to apply rule $r_i$ of ${\cal P}^{r,n}$ such that $2^{i+1} > n \geq 2^i$. We continue likewise with the resulting recursive goal until $n=0$ and we have hit the base case of the recursion.
As a consequence of this strategy,
each of the rules $r_j$ with $0 \leq j \leq i$ is applicable at most once
because from $2^{j+1} > n_j \geq 2^j$ it follows that $2^j > n_j-2^j \geq 0$.
Since at most $log_2(n)$ rules have been generated by repeated unfolding, the above claim holds.
\qed
}\end{lemma}

\subsection{Super-Linear Speedup}

With repeated unfolding up to $log_2(n)$ 
and optimal rule applications
we are replacing $n$ by $log_2(n)$ recursive calls.
For a significant speedup, we also need simplification.
To formalize the speedup, we need the following definition as a starting point.

\begin{definition}[Worst-Case Runtime Bound]
For a given recursive rule $r$, 
let $c$ be a computable unary arithmetic function
such that
for a given goal with a recursion depth $n$,
$c(n)$ is an {\em upper bound on the runtime} of the first recursive step with the original rule $r$. 
\end{definition}
Note that by this definition,
the runtime bound of the next recursive step will be $c(n{-}1)$ and so on till $c(0)$ which refers to the base case. 
We will ignore the runtime of base cases since 
their complexity is usually constant and since
they are not affected by our transformations.

A significant speedup can be achieved if we can simplify the built-in constraints in the unfolded rules so that their runtime does not double with each unfolding. 
In the best case, unfolding does not increase the runtime bound of a recursive step: instead of a time bound of $c(n)+c(n)$ in the unfolded rule, we still have a bound of $c(n)$ with simplified constraints.
We call this {\em best-case simplification}.
In the following theorem we show how this speedup can change the complexity class into a lower one.

\begin{theorem}[Super-Linear Speedup of Repeated Recursion Unfolding]\label{speeduptheo}
Given an unfolded program ${\cal P}^{r,n}$ with optimal rule applications.
We assume best-case simplification. Let the function $c(n)$ compute a time bound for the first recursive step with any recursive rule $r_i$ with $i \leq log_2(n)$ for any query with recursion depth $n$.
Then the time complexity classes\footnote{We use constant factors in the complexity classes to emphasize that the actual runtime is expected to adhere to the constant factor.}
for the original and unfolded recursion are according to Table \ref{speedup}.
\begin{figure}%[ht]
\begin{center}{\small
\begin{tabular}{|l|c|c|c|}
\cline{1-4}
Time Complexity Class $n \geq 2$ & Recursive Step $c(n)$ & Recursion $r(n)$ & Unfolded $r'(n)$\\
\cline{1-4}
(poly)logarithmic, constant $k \geq 0$ & $log_2(n)^k$ & $n log_2(n)^k$ & $log_2(n)^{k+1}$\\
\cline{1-4}
polynomial, linear $k \geq 1$ & $n^k$ & $n^{k+1}$ & $2 n^k$\\
\cline{1-4}
exponential & $2^n$ & $2^{n+1}$ & $2^{n+1}$\\
\cline{1-4}
\end{tabular}
\caption{Speedup in terms of time complexity classes with recursion length $n$}
\label{speedup}
}\end{center}
\end{figure}

\myparagraph{Proof}
Then the runtime for the recursive part of the computation $r(n)$ with $n \geq 1$ is clearly bounded
by $n c(n)$.
The time bound can be more precisely modeled by a recurrence relation\footnote{A recurrence (relation) is an equation that recursively defines a function. %sequence.
A closed-form solution of a recurrence is a finitary mathematical expression.}
of the form
$$r(n) = c(n) + r(n-1).$$

With optimal rule applications in the unfolded recursion according to Lemma \ref{optrule}, the remaining number of recursive steps to be performed is at least halved going from $n$ to $n-2^i$:
from the condition for optimal rule application
$2^{i+1} > n \geq 2^i$ it follows that $2^i > n/2$, $n-2^i < n-n/2$ and thus 
$n-2^i < n/2$.
We can therefore model the time bound by the recurrence relation
$$r'(n) = c(n) + r'(n/2).$$

Now the results in Table \ref{speedup} can be proven by solving the recurrences.
They can be verified by inserting the solutions.
For upper bounds it suffices to show that the left hand side is at least as large as the right hand side of the recurrence relation for $n \geq 2$.
For example,
let $c(n)= log_2(n)$
then 
\begin{gather*}
r(n) = n log_2(n) = log_2(n) + (n-1) log_2(n) \geq\\
log_2(n) + (n-1) log_2(n-1) = c(n) + r(n-1) \mbox{ and}\\
r'(n) = log_2(n)^2 = log_2(n) + (log_2(n)-1) log_2(n) \geq\\
log_2(n) + (log_2(n)-1)^2 = log_2(n) + log_2(n/2)^2 = c(n) + r'(n/2). \mbox{    }\qed
\end{gather*}
\end{theorem}
For linear and polynomial worst-case time complexity classes a super-linear speedup by the factor $O(n/2)$ is possible, for (poly)logarithmic or constant complexity classes a speedup of $O(n/log_2(n))$, while for exponential complexity classes no improvement of the complexity class is possible
(but the unfolded recursion will still run faster).

\section{Implementing Optimal Rule Applications}\label{secrecno}

We can readily implement optimal rule applications if we accept some overhead. 
Besides simply relying on rule order,
we present a semi-naive approach in this section called recursionless recursion.
One advantage is that this transformation completely eliminates recursion 
up to a chosen recursion depth.
It is therefore well-suited for hardware synthesis \cite{DBLP:conf/ppdp/TriossiORF12}.

\subsection{Recursionless Recursion}

We transform the recursion away completely based on the observation that in an optimal rule application, each unfolded recursive rule $r_i$ will be applied at most once only. 
We assume that rules are tried in the order in which they appear in the program as stipulated by the refined semantics of CHR.

\begin{definition}[Recursionless Recursion]
For each $i \leq log_2(n)$ we replace the rule 
$$r_i = H \simp C \, |\, D,B,R,$$
where $R$ denotes the recursive call in the body,
by the pair of rules
\begin{gather*}
r_i: H_i \simp C \, |\, D,B,R_{i-1}\\
r'_i: H^X_i \simp R^X_{i-1}
\end{gather*}
where for $H$,
$H_i$ denotes the recursive constraint $H$ whose constraint symbol $c$ has been renamed to $c_i$,
except for $H_{-1}$ which is just $H$,
and
where $H^X$ denotes a user-defined constraint whose arguments are the same sequence of distinct variables given in $X$.
The same syntax applies to $R$.
\end{definition}
Note that by this construction, each rule is applicable at most once, because there is no recursion left. For each recursion level $i$, 
relying on rule order, first rule $r_i$ will be tried.
It is either applied or otherwise rule $r'_i$ is applicable. 
In $r'_i$ we just pass the argument parameters down to the next lower rule level.
So we either we apply rule $r_i$ with no additional overhead or we apply the very simple rule $r'_i$ which has constant runtime plus the overhead of trying rule $r_i$ before.

\begin{example}[Summation, contd.]
The recursionless version of our running example is
\begin{gather*}
r_2: sum_2(N,S) \simp N>4 \, |\, S := 4*N{-}6+S1, sum_1(N{-}4,S1)\\
r'_2: sum_2(N,S) \simp %\neg N>4 \, |\, 
sum_1(N,S)\\
r_1: sum_1(N,S) \simp N>2 \, |\, S := 2*N{-}1+S1, sum_0(N{-}2,S1)\\
r'_1: sum_1(N,S) \simp %\neg N>2 \, |\, 
sum_0(N,S)\\
r_0: sum_0(N,S) \simp N>1 \, |\, S := N+S1, sum(N{-}1,S1)\\
r'_0: sum_0(N,S) \simp %\neg N>1 \, |\, 
sum(N,S)\\
sum(1,S) \simp S=1.
\end{gather*}
\end{example}

\subsection{Speedup with Worst-Case Overhead}

Rule application attempts that do not lead to rule applications cause an overhead. The overhead depends on the number of unfolded rules and the cost of trying them. In the best case, the cost for head matching attempts and guard checking is constant and the resulting overhead is neglectable. For the worst case we can assume that rule application attempts cost as much time as actual rule applications. This leads to the following lemma. It shows that even in the worst case super-linear speedup is still obtained for most recursion depths.

\begin{lemma}[Speedup with Worst-Case Overhead]\label{overheadlem}
Let $N > n$.
Assume we have repeatedly unfolded for recursion depth $N$, i.e. up to $log_2(N)$.
We have a query with recursion depth $n$ and complexities $c(n)$, $r(n)$ and $r'(N)$ as calculated in Theorem \ref{speeduptheo}.
Then Table \ref{overhead} shows the upper bounds on the worst-case overhead for the corresponding complexity classes.
\begin{figure}%[htb]
\begin{center}{\small
\begin{tabular}{|l|c|c|c|c|}
\cline{1-5}
Complexity $N > n \geq 2$ & Recursion & Unfolded & Just Rule Order & Recursionless\\
\cline{1-5}
(poly)logarithmic, c. $k \geq 0$ & $n log_2(n)^k$ & $log_2(n)^{k+1}$ & $log_2(N) log_2(n)^{k+1}$ & $2 log_2(N) log_2(n)^{k}$\\
\cline{1-5}
polynomial, linear $k \geq 1$ & $n^{k+1}$ & $2 n^k$ & $2 log_2(N) n^k$ & $(log_2(N)+2)  n^k$\\
\cline{1-5}
exponential & $2^{n+1}$ & $2^{n+1}$ & $log_2(N) 2^{n+1}$ & $(log_2(N)+2) 2^{n}$\\
\cline{1-5}
\end{tabular}
\caption{Speedup with worst-case overhead when unfolded up to $log_2(N)$}
\label{overhead}
}\end{center}
\end{figure}

\myparagraph{Proof}
When we use rule order alone, up to $log_2(N)$ rules will be tried in vain before a rule will be applied. This happens for each recursive call. For the first recursive call this incurs a runtime of $c(n)$ with each attempt in the worst case or each application, for the second of $c(n/2)$ and so on. We already know that the sum of these runtimes is $r'(n)$. So instead of the runtime $r'(n)$ we have a worst-case time bound of
$$log_2(N) r'(n).$$

With recursionless recursion we try each rule at most once. 
With rule order the handling of the additional rules $r'_i$ at each recursion level takes constant time and we can ignore them in the following.
First we try $log_2(N)-log_2(n)$ unfolded rules $r_i$ in vain before we reach the rule corresponding to level $log_2(n)$. From then on, the worst case is the application of all rules below.
So overall,
$log_2(N)-log_2(n)$ rules are tried once each with runtime $c(n)$ and then a runtime of $r'(n)$ is incurred, resulting in a bound
$$(log_2(N)-log_2(n)) c(n) + r'(n).$$
Using these formulas with the complexities for $c(n)$, $r(n)$ and $r'(N)$ from Theorem \ref{speeduptheo} gives the upper bounds that are listed in somewhat simplified form in Table \ref{overhead}.
\qed
\end{lemma}
From Table \ref{overhead} we can see that the worst case overhead is a factor of $O(log_2(N))$ for rule order alone.
For recursionless recursion, it is a worst case factor of $O(2 log_2(N)/log_2(n))$ for polylogarithmic complexity classes, and a factor of $O(log_2(N)/2)$ for polynomial and exponential complexity.

In this worst case of recursionless recursion, all complexity classes but exponential complexity are still lower than the complexity classes of the original recursions as long as $n$ is larger than $log_2(N)$.

\myparagraph{Unbounded Recursion}
With recursionless recursion, the recursion depth is bounded by $N$. We can simply eliminate this bound by re-introducing recursion for the most unfolded rule $r_k$ with $k = \lfloor log_2(N)\rfloor$:
$$r_k = H_k \simp C \, |\, D,B,R_k.$$
For $N \leq n$ we now apply rule $r_k$ as long as possible.
Of course, recursionless recursion is now a misnomer.

\section{Examples with Benchmarks}\label{secbench}

We have implemented a simple prototype for performing repeated recursion unfolding in CHR and Prolog relying on rule order and recursionless recursion.
The source code of the resulting example programs is listed in the appendix.
In our experiments, we used the CHR system in SWI Prolog Version 6.2.1 running on an
Apple Mac mini 2018 with Intel Core i5 8GB RAM and OS-X 10.14.6.
In our benchmarks, unfolding is up to the level given by $log_2(N)$ with $N>n$.

\subsection{Benchmarks for Summation Example}

All recursive rules of $sum$ have a constant time for matching the head, checking the guard, computing the sum and doing the recursive call including the subtraction.
The recursion depth is determined by the input number $n$. 

\myparagraph{Super-Linear Speedup}
Table \ref{benchsum} lists our benchmark results for repeated unfolding up to $log_2(N)=25$ and input numbers $n$ from 1024 to 8192.
Times are in milliseconds. Summation has been performed once for all numbers in the given range. The given timings are the sum of these execution times.
The unfolded versions showed constant timings for the number ranges over the complete set of numbers, with recursionless recursion performing about 200 summations per millisecond.
\begin{figure}%[ht]
\begin{center}{\small
\begin{tabular}{|l|r|r|r|r|r|r|r|r|}
\cline{1-8}
\multicolumn{8}{|c|}{Summation $log_2(N)=25$}\\
\cline{1-8}
Input $n$ %& 1000x1 
& 1024-2047 & 2048-3071 & 3072-4095 & 4096-5119 & 5120-6143 & 6144-7167 & 7168-8192\\
\cline{1-8}
Original %& 0.34 
& 240 & 481 & 839 & 1115 & 1507 & 1832 & 2258\\
Rule Order %& 1.98 
& 19 & 20 & 20 & 21 & 20 & 22 & 21\\
Rec.less %& 2.44 
& 4 & 5 & 4 & 5 & 5 & 5 & 5\\
\cline{1-8}
Prolog %& 0.24 
& 185 & 369 & 663 & 873 & 1207 & 1457 & 1816\\
\cline{1-8}
\end{tabular}
\caption{Benchmarks of Summation}
\label{benchsum}
}\end{center}
\end{figure}
The row {\em Original} shows the linear runtime of the given recursive summation.
Row {\em Rule Order} refers to the unfolded recursion of $sum$ relying on rule order alone, which already shows constant time and at least a ten-fold speedup. {\em Recursionless} refers to the recursionless recursion, it is a factor of four faster than the unfolded recursion, again showing constant time. 
By comparison, the Prolog implementation of the original rules using the cut operator for efficiency shows again the linear time behavior.

According to the Speedup Theorem \ref{speeduptheo},
for the original recursion we expect linear time complexity $O(n)$. 
According to Lemma \ref{overheadlem},
for the unfolded program with rule order the complexity is
$O(log_2(N) log_2(n))$ and with recursionless recursion $O(log_2(N))$.
Since $N$ is fixed in our benchmarks, this amounts to constant time.

We attribute the better-than-estimated constant time behavior of the unfolded recursion with rule order to the strong variation in rule attempts and applications with subsequent numbers. For example, while the number $1025$ requires just one recursive step with rule $r_{10}$, the previous number $1024$ needs all smaller recursive steps from $r_9$ down to $r_0$. 
This behavior leads to constant runtime when we sum over ranges of numbers as in Table \ref{benchsum}.

\subsection{Complete Example List Reversal}

This classical program reverses a given list in a naive way.
The constraint $r(A,B)$ holds if list $B$ is the reversal of list $A$.
We use Prolog notation for lists.
\begin{gather*}
r([C|A], D) \simp r(A, B), a(B, [C], D)\\
r([], D) \simp D=[].
\end{gather*}
The built-in constraint $a(X,Y,Z)$ appends (concatenates) two lists $X$ and $Y$ into a third list $Z$. Its runtime is linear in the length of the first list.

\myparagraph{Repeated Recursion Unfolding}
We try to unfold the recursive rule with a copy of itself:
\begin{gather*}
r([C|A], D) \simp r(A, B), a(B, [C], D)\\
r([C'|A'], D') \simp r(A', B'), a(B', [C'], D').
\end{gather*}
However the recursive call in the original rule $r(A, B)$ is not an instance of the head $r([C'|A'], D')$ of the copy of the rule.
This is a mere technicality.
We move $[C'|A']$ into the guard and replace it by a new variable $A''$:
$$r(A'', D') \simp A''=[C'|A'] \, |\, r(A', B'), a(B', [C'], D').$$
This does not change the semantics of the rule because
$$r([C'|A'], D') \equiv_{\{C',A',D'\}} r(A'', D'), A''=[C'|A'].$$
The unfolding is now correct with $A''$ substituted by $A$ in the guard
\begin{gather*}
r([C|A], D){\simp}A{=}[C'|A'] \, | \,
r(A, B){=}r(A'', D'), r(A', B'),\\
 a(B', [C'], D'), a(B, [C], D).
\end{gather*}
Next we use the guard equality $A=[C'|A']$ to substitute the variable $A$ in the head $r([C|A], D)$ back to $[C'|A']$.
This is correct since
$$r([C|A], D), A=[C'|A'] \equiv_{\{C,A,D,C',A'\}} r([C,C'|A']], D), A=[C'|A'].$$
Now we proceed with the simplification for unfolded rules as defined.
For the head and guard we simplify
$$r([C,C'|A'], D), A=[C'|A'] \equiv_{\{C,D,C',A'\}} r([C, C'|A'], D).$$
For the body we simplify
\begin{gather*}
A{=}[C'|A'], r(A, B){=}r(A'', D'), r(A', B'), a(B', [C'], D'), a(B, [C], D)\!\equiv_{\{C,D,C',A'\}}\\
r(A', B'), a(B', [C'], B), a(B, [C], D)
\equiv_{\{C,D,C',A'\}}\\
r(A', B'), a(B', [C',C], D).
\end{gather*}
The eureka moment for best-case simplification is that
we can merge the two calls to constraint $a$ into one if we concatenate their known second arguments.

As a result the unfolded rule is simplified into the rule
$$r([C, C'|A'], D) \simp r(A', B'), a(B', [C', C], D).$$

Repeated recursion unfolding thrice results in the following simplified rules
\begin{gather*}
r([J, I, H, G, F, E, D, C|A], K) \simp r(A, B), a(B, [C, D, E, F, G, H, I, J], K)\\
r([F, E, D, C|A], G) \simp r(A, B), a(B, [C, D, E, F], G)\\
r([D, C|A], E) \simp r(A, B), a(B, [C, D], E)\\
r([C|A], D) \simp r(A, B), a(B, [C], D)\\
r([], D) \simp D=[].
\end{gather*}
Note that we see here a worst case of program size increase. With each unfolding, the rule size doubles. Still, this does not lead to code explosion since we unfold only up to $log(n)$ times and therefore the size of {\em all} rules added is proportional to $n$.

\myparagraph{Super-Linear Speedup}
Repeated recursion unfolding was performed up to $log_2(N)=10$. The benchmarks in Table \ref{benchrev} cover all possible list lengths from 128 to 1929.
Times are in milliseconds. They give the sum of runtimes for each of the 256 list lengths in the given range of numbers. 
\begin{figure}%[ht]
\begin{center}{\small
\begin{tabular}{|l|r|r|r|r|r|r|r|}
\cline{1-8}
\multicolumn{8}{|c|}{List Reversal $log_2(N)=10$}\\
\cline{1-8}
List Lengths & 128-383 & 384-639 & 640-895 & 896-1151 & 1152-1407 & 1408-1663 & 1664-1929\\
\cline{1-8}
Original & 496.0 & 1847.0 & 4442.0 & 8630.0 & 14696.0 & 23445.0 & 35487.0\\
\cline{1-8}
Hand-Optimized & 7.6 & 15.4 & 22.8 & 30.2 & 37.4 & 45.1 & 52.6\\
Rule Order & 5.7 & 8.0 & 11.0 & 12.4 & 11.8 & 16.8 & 22.1\\
Recursionless & 3.6 & 5.8 & 8.6 & 9.3 & 9.1 & 13.2 & 17.7\\
\cline{1-8}
Prolog Built-In & 4.2 & 8.6 & 12.7 & 16.6 & 20.9 & 24.7 & 29.3\\
\cline{1-8}
\end{tabular}
\caption{Benchmarks of List Reversal}
\label{benchrev}
}\end{center}
\end{figure}
The row {\em Original} stands for the original recursion implementing naive list reversal.
Row {\em Hand-Optimized} stands for the linear time non-naive list reversal written in CHR.
Row {\em Rule Order} stands for the repeated unfolding of naive list reversal relying on rule order alone.
Row {\em Recursionless} stands for the repeated unfolding of naive list reversal were recursion is completely eliminated.
Row {\em Prolog Built-In} stands for the built-in list reversal of SWI Prolog.

According to our Speedup Theorem \ref{speeduptheo} and Lemma \ref{overheadlem}, we expect quadratic behavior for the original recursion and linear behavior for the unfolded recursion, were the recursionless recursion should be twice as fast. This actually holds as the benchmarks show.

All timings show linear time behavior except the original list reversal without unfolding.
The fastest program version with recursionless recursion reverses about 100 list elements per millisecond.
The transformed code executed twice as fast as the hand-coded well-known linear version of list reversal that does not use list concatenation. 
Surprisingly, our recursionless recursion proved even faster than the built-in optimized {\tt reverse/2} of SWI Prolog.

Note that the runtimes around $1024$ almost stay constant. Our most unfolded rule with $i=N=10$ handles a list with $2^{10} = 1024$ elements.
Again we attribute this averaging effect of considering ranges of inputs.
For example, a list of length $1024$ needs one rule application with rule $r_{10}$ at recursion level $i=10$ before the base case is reached, while with $1023$ we need to apply all rules for each level $i<10$.

\section{Related Work}\label{discuss}

Program transformation to improve efficiency is usually concerned with a strategy for the combination of unfolding and folding techniques to replace code starting with \cite{burstall1977transformation}. In the literature, recursion is usually unfolded several times together with the base case and then simplified.
We rely solely on unfolding the recursive part and repeat the process. 
We ignore the base case.
We add redundant rules this way but never remove any.

Program transformation as a tool for specific aims and 
applications is abundant in CHR, for an overview see
\cite{chr_survey_tplp10,DBLP:conf/ruleml/Fruhwirth15}. 
General methods for
unfolding \cite{gabbrielli2015unfolding} exist (which we have used in this paper), for specializing rules with regard to a specific given query \cite{fru_specialization_lopstr04}, 
and for optimization induced by confluence \cite{abd_fru_integration_lopstr03}.

In \cite{Hansen:1991:PUM:115865.115892} {\em unfolding-based meta-level systems} for Prolog consist of a hierarchy of additional meta-rules and a hierarchical execution scheme, i.e. a change in semantics. These rules are described as shortcuts obtained from unfold/fold operations. Presumably, 
when adapted to Prolog, the rules generated by repeated recursion unfolding could be considered as meta-rules in this generic framework. 

Directly related literature is sparse. One reason could be that our approach is based on generating and keeping redundant rules which seems counter-intuitive at first.
Moreover, 
super-linear speedups are a rare feat and mostly concern parallel programs, while our approach applies to sequential programs.
An exception is work based on supercompilation. In this program transformation,
generalisation while unfolding increases the chance for folding.
Using advanced generalisation techniques for supercompilation, distillation \cite{hamilton09}
and equality indices \cite{glueck16} can achieve super-linear speedup on some examples.
In contrast, our approach is straightforward, as it does not involve generalisation or folding, and is applied to a programming language in practical use.

\section{Conclusions and Future Work}

Repeated recursion unfolding is a new approach that repeatedly
unfolds a recursion with itself and simplifies it while keeping all unfolded rules. We proved its correctness.

We showed there exists an optimal rule application strategy such that
significant speedups are possible.
We proved a super-linear speedup theorem in case of best-case simplification.
Then our program transformation lowers the time complexity of a recursion
for polylogarithmic classes by a factor of $O(n/log_2(n))$ and for polynomial classes by a factor of $O(n)$.

For a semi-naive implementation of repeated recursion unfolding, recursionless recursion, we proved an overhead which is linear in the number of unfolded rules. 
Super-linear speedup is still obtained in most cases.
The actual runtime improvement quickly reaches several orders of magnitude.
Our approach improves the time complexity from linear to constant for the summation example and from quadratic to linear for list reversal.
The latter runs faster than a built-in hand-optimized version.

While our speedups are super-linear, it is too early to tell which recursive algorithms allow for the necessary best-case simplification.
A good simplification requires some insight and thus in general cannot be found automatically.

We defined and implemented repeated recursion unfolding using the rule-based language CHR (Constraint Handling Rules), but we think our approach can be applied to mainstream programming languages and hardware synthesis as well.

\myparagraph{Ongoing and future work}
To extend our approach to mutual recursion as well as multiple recursive rules should be straightforward.
Indeed, we already could derive a novel double recursive linear-time algorithm from a naive exponential-time double recursion for Fibonacci numbers.

In this paper, the super-linear speedup is bounded in the sense that it holds up to a chosen upper bound on the number of recursive steps. 
In ongoing work, we have succeeded for the examples in this paper to extend our approach to 
run-time dynamic on-the-fly just-in-time %!!!!!!
repeated recursion unfolding so that the super-linear speedup is unbounded, i.e. independent of the recursion depth.

The implementation overhead of optimal rule applications can be reduced with indexing on the recursion depth.
We already have implemented indexing for the examples in this paper.

One could also transfer our approach from recursion to loop constructs in other programming languages.
We have already done examples for Java.

\medskip
{\bf Acknowledgements.} 
Part of this research work was performed during the sabbatical of the author in summer semester 2020.
We thank the anonymous reviewers for their skepticism which helped to clarify the contribution of the paper.

\bibliographystyle{alpha} %! abbrv, alpha is longer
\bibliography{recunfold,biblio,chrjust}

\appendix

\section{Proof of Correctness of Rule Simplification}

For correctness of rule simplification (see Subsection \ref{rulesimp})
we have to show that the same transitions $S \mapsto T$ are possible with rule $r$ and rule $\mathit{simplify}(r)$. Here is the full proof for Theorem 1.

\newtheorem{theorem*}{Theorem}%[Section] %!!!! to get Theorem 1 again

\begin{theorem*}[Correctness of Rule Simplification]
{%\rm
Let $r = (H \Leftrightarrow C \, |\, D \land B)$ be a rule and
let $s = (H' \Leftrightarrow C' \, |\, D' {\setminus} C' \land B')$ be the simplified rule $\mathit{simplify}(r)$.
For any state $S$ and variables ${\cal V}$, $S \mapsto_r T$ iff $S \mapsto_s T$.

\myparagraph{Proof}
According to the definition of a CHR transition and $\mathit{simplify}(r)$, we know that
\begin{center}
$S \mapsto_r T \mbox{ iff }
S \equiv_{\cal V} (H \land C \land G)  \not\equiv \false \mbox{ and } (C \land D \land B \land G) \equiv_{\cal V} T$\\
$S \mapsto_s T \mbox{ iff }
S \equiv_{\cal V} (H' \land C' \land G')  \not\equiv \false \mbox{ and } (C' \land D' {\setminus} C' \land B' \land G') \equiv_{\cal V} T$\\
$(H \land C) \equiv_{\cal V'} (H' \land C')$\\
$(C \land D \land B) \equiv_{\cal V'} (D' \land B').$
\end{center}
It suffices to show that $S \mapsto_r T$ implies $S \mapsto_s T$, since the implication in the other direction is symmetric and can be shown in the same way.
Hence we have to show that there exists a goal $G'$ such that
\begin{center}
$S \equiv (H \land C \land G) \equiv_{\cal V} (H' \land C' \land G') \mbox{ if }
(H \land C) \equiv_{\cal V'} (H' \land C')$ and\\
$T \equiv (C \land D \land B \land G) \equiv_{\cal V} (C' \land D' {\setminus} C' \land B' \land G')  \mbox{ if } (C \land D \land B) \equiv_{\cal V'} (D' \land B').$
\end{center}
To show the first part concerning state $S$, we proceed as follows.
We choose $G' = C \land G$. %!!!
Let $G=(G_C \land G_B)$ where $G_C$ are the built-in constraints and $G_B$ are the user-defined constraints of $G$.
According to the definition of state equivalence and its transitivity
we have to show that the two states equivalent to $S$ are equivalent
\begin{center}
$\CT \models 
\forall (C \land G_C \rightarrow \exists \bar y ((H \land G_B) = (H' \land G_B) \land C' \land C \land G_C))$\\      
$\land \ \ \forall (C' \land C \land G_C \rightarrow \exists \bar x ((H \land G_B) = (H' \land G_B) \land C \land G_C))$\\
if\\
$\CT \models 
\forall (C \rightarrow \exists \bar y' ((H = H') \land C'))$\\      
$\land \ \ \forall (C' \rightarrow \exists \bar x' ((H = H') \land C)),$
\end{center}
where the $\bar x',\bar y',\bar x,\bar y$ are local variables, 
such that
\begin{itemize}
\item $\bar x'$ are the variables in $H \land C$ only that are not in ${\cal V'}$,
\item $\bar y'$ are the variables in $H' \land C'$ only that are not in ${\cal V'}$,
\item $\bar x$ are the variables in $H \land C \land G$ that are not in ${\cal V}$,
\item $\bar y$ are the variables in $H' \land C' \land G'$ that are not in ${\cal V}$.
\end{itemize}
By the definition of simplification, ${\cal V'}$ are the variables in $H$ and $H'$.
By definition of the transition, 
$\bar x$ includes the local variables of the given rule, i.e. those not in the head $H$,
and $\bar y$ includes the local variables of the simplified rule, i.e. those not in the head $H'$.

From these facts we can conclude that $\bar x$ includes the variables of $\bar x'$ and that $\bar y$ includes the variables of $\bar y'$.
The remaining variables in $\bar x$ are those in $G$ that are not in ${\cal V}$,
the remaining variables in $\bar y$ are those in $G'$ that are not in ${\cal V}$.
Since $G' = C \land G$, $\bar y$ includes the variables of $\bar x$ in addition those in $C'$ that are not in $H'$.
We can therefore write $\bar x$ as $\bar x'\bar z$ and
$\bar y$ as $\bar y'\bar z\bar w$.

Now consider the precondition

\begin{center}
$\CT \models \forall (C \rightarrow \exists \bar y' ((H = H') \land C'))$\\      
$\land \ \ \forall (C' \rightarrow \exists \bar x' ((H = H') \land C)),$
\end{center}
We can always add existential quantifiers to the conclusion of an implication, since $A \rightarrow B$ implies $A \rightarrow \exists \bar v B$ in first-order predicate logic.
\begin{center}
$\CT \models \forall (C \rightarrow \exists \bar y'\bar z \bar w ((H = H') \land C'))$\\      
$\land \ \ \forall (C' \rightarrow \exists \bar x'\bar z ((H = H') \land C)).$
\end{center}
Since $\CT \models G_B = G_B$ we can extend $H = H'$ to $(H \land G_B) = (H' \land G_B)$
and since
$A \rightarrow \exists \bar v A$ for all $A$ and $\bar v$ we can extend the implications on both sides by $G_C$ and $C \land G_C$, respectively.
Since $\bar x = \bar x'\bar z$ and
$\bar y = \bar y'\bar z\bar w$,
we finally arrive at
\begin{center}
$\CT \models 
\forall (C \land G_C \rightarrow \exists \bar y ((H \land G_B) = (H' \land G_B) \land C' \land C \land G_C))$\\      
$\land \ \ \forall (C' \land C \land G_C \rightarrow \exists \bar x ((H \land G_B) = (H' \land G_B) \land C \land G_C)).$\\
\end{center}

In the same manner we can show that the two states equivalent to state $T$ are equivalent by observing that the local variables of $C,D,B$ and $C',D',B'$
are the local variables of the respective rules and are thus also included in the local variables of the two states equivalent to $T$.
Also note that $(C' \land D' {\setminus} C') = (C' \land D')$.
\qed
}%rm
\end {theorem*}

\section{Program Code Listings}

\subsection{Summation, unfolded up to N=$2^{25}$}

{\small 
\begin{verbatim}
sum25(A,C) <=> A>33554432 | B is A-33554432, sum24(B, D), 
                  C is 33554432*A-562949936644096+D.
sum25(A,B) <=> sum24(A,B).
sum24(A,C) <=> A>16777216 | B is A-16777216, sum23(B, D), 
                  C is 16777216*A-140737479966720+D.
sum24(A,B) <=> sum23(A,B).
sum23(A,C) <=> A>8388608 | B is A-8388608, sum22(B, D), 
                  C is 8388608*A-35184367894528+D.
sum23(A,B) <=> sum22(A,B).
sum22(A,C) <=> A>4194304 | B is A-4194304, sum21(B, D), 
                  C is 4194304*A-8796090925056+D.
sum22(A,B) <=> sum21(A,B).
sum21(A,C) <=> A>2097152 | B is A-2097152, sum20(B, D), 
                  C is 2097152*A-2199022206976+D.
sum21(A,B) <=> sum20(A,B).
sum20(A,C) <=> A>1048576 | B is A-1048576, sum19(B, D), 
                  C is 1048576*A-549755289600+D.
sum20(A,B) <=> sum19(A,B).
sum19(A,C) <=> A>524288 | B is A-524288, sum18(B, D), 
                  C is 524288*A-137438691328+D.
sum19(A,B) <=> sum18(A,B).
sum18(A,C) <=> A>262144 | B is A-262144, sum17(B, D), 
                  C is 262144*A-34359607296+D.
sum18(A,B) <=> sum17(A,B).
sum17(A,C) <=> A>131072 | B is A-131072, sum16(B, D), 
                  C is 131072*A-8589869056+D.
sum17(A,B) <=> sum16(A,B).
sum16(A,C) <=> A>65536 | B is A-65536, sum15(B, D), 
                  C is 65536*A-2147450880+D.
sum16(A,B) <=> sum15(A,B).
sum15(A,C) <=> A>32768 | B is A-32768, sum14(B, D), 
                  C is 32768*A-536854528+D.
sum15(A,B) <=> sum14(A,B).
sum14(A,C) <=> A>16384 | B is A-16384, sum13(B, D), 
                  C is 16384*A-134209536+D.
sum14(A,B) <=> sum13(A,B).
sum13(A,C) <=> A>8192 | B is A-8192, sum12(B, D), C is 8192*A-33550336+D.
sum13(A,B) <=> sum12(A,B).
sum12(A,C) <=> A>4096 | B is A-4096, sum11(B, D), C is 4096*A-8386560+D.
sum12(A,B) <=> sum11(A,B).
sum11(A,C) <=> A>2048 | B is A-2048, sum10(B, D), C is 2048*A-2096128+D.
sum11(A,B) <=> sum10(A,B).
sum10(A,C) <=> A>1024 | B is A-1024, sum9(B, D), C is 1024*A-523776+D.
sum10(A,B) <=> sum9(A,B).
sum9(A,C) <=> A>512 | B is A-512, sum8(B, D), C is 512*A-130816+D.
sum9(A,B) <=> sum8(A,B).
sum8(A,C) <=> A>256 | B is A-256, sum7(B, D), C is 256*A-32640+D.
sum8(A,B) <=> sum7(A,B).
sum7(A,C) <=> A>128 | B is A-128, sum6(B, D), C is 128*A-8128+D.
sum7(A,B) <=> sum6(A,B).
sum6(A,C) <=> A>64 | B is A-64, sum5(B, D), C is 64*A-2016+D.
sum6(A,B) <=> sum5(A,B).
sum5(A,C) <=> A>32 | B is A-32, sum4(B, D), C is 32*A-496+D.
sum5(A,B) <=> sum4(A,B).
sum4(A,C) <=> A>16 | B is A-16, sum3(B, D), C is 16*A-120+D.
sum4(A,B) <=> sum3(A,B).
sum3(A,C) <=> A>8 | B is A-8, sum2(B, D), C is 8*A-28+D.
sum3(A,B) <=> sum2(A,B).
sum2(A,C) <=> A>4 | B is A-4, sum1(B, D), C is 4*A-6+D.
sum2(A,B) <=> sum1(A,B).
sum1(A,C) <=> A>2 | B is A-2, sum0(B, D), C is 2*A-1+D.
sum1(A,B) <=> sum0(A,B).
sum0(A,C) <=> A>1 | B is A-1, sum(B, D), C is A+D.
sum0(A,B) <=> sum(A,B).
sum(1, A) <=> A=1.
\end{verbatim}
} %small

\subsection{List Reversal, unfolded up to N=$2^{10}$}

{\small
\begin{verbatim}
r10([L39,K39,J39,I39,H39,G39,F39,E39,D39,C39,B39,A39,Z38,Y38,X38,W38,V38,U38,T38,S38,R38,Q38,P38,O38,N38,M38,L38,K38,J38,I38,H38,G38,F38,E38,D38,C38,B38,A38,Z37,Y37,X37,W37,V37,U37,T37,S37,R37,Q37,P37,O37,N37,M37,L37,K37,J37,I37,H37,G37,F37,E37,D37,C37,B37,A37,Z36,Y36,X36,W36,V36,U36,T36,S36,R36,Q36,P36,O36,N36,M36,L36,K36,J36,I36,H36,G36,F36,E36,D36,C36,B36,A36,Z35,Y35,X35,W35,V35,U35,T35,S35,R35,Q35,P35,O35,N35,M35,L35,K35,J35,I35,H35,G35,F35,E35,D35,C35,B35,A35,Z34,Y34,X34,W34,V34,U34,T34,S34,R34,Q34,P34,O34,N34,M34,L34,K34,J34,I34,H34,G34,F34,E34,D34,C34,B34,A34,Z33,Y33,X33,W33,V33,U33,T33,S33,R33,Q33,P33,O33,N33,M33,L33,K33,J33,I33,H33,G33,F33,E33,D33,C33,B33,A33,Z32,Y32,X32,W32,V32,U32,T32,S32,R32,Q32,P32,O32,N32,M32,L32,K32,J32,I32,H32,G32,F32,E32,D32,C32,B32,A32,Z31,Y31,X31,W31,V31,U31,T31,S31,R31,Q31,P31,O31,N31,M31,L31,K31,J31,I31,H31,G31,F31,E31,D31,C31,B31,A31,Z30,Y30,X30,W30,V30,U30,T30,S30,R30,Q30,P30,O30,N30,M30,L30,K30,J30,I30,H30,G30,F30,E30,D30,C30,B30,A30,Z29,Y29,X29,W29,V29,U29,T29,S29,R29,Q29,P29,O29,N29,M29,L29,K29,J29,I29,H29,G29,F29,E29,D29,C29,B29,A29,Z28,Y28,X28,W28,V28,U28,T28,S28,R28,Q28,P28,O28,N28,M28,L28,K28,J28,I28,H28,G28,F28,E28,D28,C28,B28,A28,Z27,Y27,X27,W27,V27,U27,T27,S27,R27,Q27,P27,O27,N27,M27,L27,K27,J27,I27,H27,G27,F27,E27,D27,C27,B27,A27,Z26,Y26,X26,W26,V26,U26,T26,S26,R26,Q26,P26,O26,N26,M26,L26,K26,J26,I26,H26,G26,F26,E26,D26,C26,B26,A26,Z25,Y25,X25,W25,V25,U25,T25,S25,R25,Q25,P25,O25,N25,M25,L25,K25,J25,I25,H25,G25,F25,E25,D25,C25,B25,A25,Z24,Y24,X24,W24,V24,U24,T24,S24,R24,Q24,P24,O24,N24,M24,L24,K24,J24,I24,H24,G24,F24,E24,D24,C24,B24,A24,Z23,Y23,X23,W23,V23,U23,T23,S23,R23,Q23,P23,O23,N23,M23,L23,K23,J23,I23,H23,G23,F23,E23,D23,C23,B23,A23,Z22,Y22,X22,W22,V22,U22,T22,S22,R22,Q22,P22,O22,N22,M22,L22,K22,J22,I22,H22,G22,F22,E22,D22,C22,B22,A22,Z21,Y21,X21,W21,V21,U21,T21,S21,R21,Q21,P21,O21,N21,M21,L21,K21,J21,I21,H21,G21,F21,E21,D21,C21,B21,A21,Z20,Y20,X20,W20,V20,U20,T20,S20,R20,Q20,P20,O20,N20,M20,L20,K20,J20,I20,H20,G20,F20,E20,D20,C20,B20,A20,Z19,Y19,X19,W19,V19,U19,T19,S19,R19,Q19,P19,O19,N19,M19,L19,K19,J19,I19,H19,G19,F19,E19,D19,C19,B19,A19,Z18,Y18,X18,W18,V18,U18,T18,S18,R18,Q18,P18,O18,N18,M18,L18,K18,J18,I18,H18,G18,F18,E18,D18,C18,B18,A18,Z17,Y17,X17,W17,V17,U17,T17,S17,R17,Q17,P17,O17,N17,M17,L17,K17,J17,I17,H17,G17,F17,E17,D17,C17,B17,A17,Z16,Y16,X16,W16,V16,U16,T16,S16,R16,Q16,P16,O16,N16,M16,L16,K16,J16,I16,H16,G16,F16,E16,D16,C16,B16,A16,Z15,Y15,X15,W15,V15,U15,T15,S15,R15,Q15,P15,O15,N15,M15,L15,K15,J15,I15,H15,G15,F15,E15,D15,C15,B15,A15,Z14,Y14,X14,W14,V14,U14,T14,S14,R14,Q14,P14,O14,N14,M14,L14,K14,J14,I14,H14,G14,F14,E14,D14,C14,B14,A14,Z13,Y13,X13,W13,V13,U13,T13,S13,R13,Q13,P13,O13,N13,M13,L13,K13,J13,I13,H13,G13,F13,E13,D13,C13,B13,A13,Z12,Y12,X12,W12,V12,U12,T12,S12,R12,Q12,P12,O12,N12,M12,L12,K12,J12,I12,H12,G12,F12,E12,D12,C12,B12,A12,Z11,Y11,X11,W11,V11,U11,T11,S11,R11,Q11,P11,O11,N11,M11,L11,K11,J11,I11,H11,G11,F11,E11,D11,C11,B11,A11,Z10,Y10,X10,W10,V10,U10,T10,S10,R10,Q10,P10,O10,N10,M10,L10,K10,J10,I10,H10,G10,F10,E10,D10,C10,B10,A10,Z9,Y9,X9,W9,V9,U9,T9,S9,R9,Q9,P9,O9,N9,M9,L9,K9,J9,I9,H9,G9,F9,E9,D9,C9,B9,A9,Z8,Y8,X8,W8,V8,U8,T8,S8,R8,Q8,P8,O8,N8,M8,L8,K8,J8,I8,H8,G8,F8,E8,D8,C8,B8,A8,Z7,Y7,X7,W7,V7,U7,T7,S7,R7,Q7,P7,O7,N7,M7,L7,K7,J7,I7,H7,G7,F7,E7,D7,C7,B7,A7,Z6,Y6,X6,W6,V6,U6,T6,S6,R6,Q6,P6,O6,N6,M6,L6,K6,J6,I6,H6,G6,F6,E6,D6,C6,B6,A6,Z5,Y5,X5,W5,V5,U5,T5,S5,R5,Q5,P5,O5,N5,M5,L5,K5,J5,I5,H5,G5,F5,E5,D5,C5,B5,A5,Z4,Y4,X4,W4,V4,U4,T4,S4,R4,Q4,P4,O4,N4,M4,L4,K4,J4,I4,H4,G4,F4,E4,D4,C4,B4,A4,Z3,Y3,X3,W3,V3,U3,T3,S3,R3,Q3,P3,O3,N3,M3,L3,K3,J3,I3,H3,G3,F3,E3,D3,C3,B3,A3,Z2,Y2,X2,W2,V2,U2,T2,S2,R2,Q2,P2,O2,N2,M2,L2,K2,J2,I2,H2,G2,F2,E2,D2,C2,B2,A2,Z1,Y1,X1,W1,V1,U1,T1,S1,R1,Q1,P1,O1,N1,M1,L1,K1,J1,I1,H1,G1,F1,E1,D1,C1,B1,A1,Z,Y,X,W,V,U,T,S,R,Q,P,O,N,M,L,K,J,I,H,G,F,E,D,C|A],M39) <=> r9(A,B),
        append(B,[C,D,E,F,G,H,I,J,K,L,M,N,O,P,Q,R,S,T,U,V,W,X,Y,Z,A1,B1,C1,D1,E1,F1,G1,H1,I1,J1,K1,L1,M1,N1,O1,P1,Q1,R1,S1,T1,U1,V1,W1,X1,Y1,Z1,A2,B2,C2,D2,E2,F2,G2,H2,I2,J2,K2,L2,M2,N2,O2,P2,Q2,R2,S2,T2,U2,V2,W2,X2,Y2,Z2,A3,B3,C3,D3,E3,F3,G3,H3,I3,J3,K3,L3,M3,N3,O3,P3,Q3,R3,S3,T3,U3,V3,W3,X3,Y3,Z3,A4,B4,C4,D4,E4,F4,G4,H4,I4,J4,K4,L4,M4,N4,O4,P4,Q4,R4,S4,T4,U4,V4,W4,X4,Y4,Z4,A5,B5,C5,D5,E5,F5,G5,H5,I5,J5,K5,L5,M5,N5,O5,P5,Q5,R5,S5,T5,U5,V5,W5,X5,Y5,Z5,A6,B6,C6,D6,E6,F6,G6,H6,I6,J6,K6,L6,M6,N6,O6,P6,Q6,R6,S6,T6,U6,V6,W6,X6,Y6,Z6,A7,B7,C7,D7,E7,F7,G7,H7,I7,J7,K7,L7,M7,N7,O7,P7,Q7,R7,S7,T7,U7,V7,W7,X7,Y7,Z7,A8,B8,C8,D8,E8,F8,G8,H8,I8,J8,K8,L8,M8,N8,O8,P8,Q8,R8,S8,T8,U8,V8,W8,X8,Y8,Z8,A9,B9,C9,D9,E9,F9,G9,H9,I9,J9,K9,L9,M9,N9,O9,P9,Q9,R9,S9,T9,U9,V9,W9,X9,Y9,Z9,A10,B10,C10,D10,E10,F10,G10,H10,I10,J10,K10,L10,M10,N10,O10,P10,Q10,R10,S10,T10,U10,V10,W10,X10,Y10,Z10,A11,B11,C11,D11,E11,F11,G11,H11,I11,J11,K11,L11,M11,N11,O11,P11,Q11,R11,S11,T11,U11,V11,W11,X11,Y11,Z11,A12,B12,C12,D12,E12,F12,G12,H12,I12,J12,K12,L12,M12,N12,O12,P12,Q12,R12,S12,T12,U12,V12,W12,X12,Y12,Z12,A13,B13,C13,D13,E13,F13,G13,H13,I13,J13,K13,L13,M13,N13,O13,P13,Q13,R13,S13,T13,U13,V13,W13,X13,Y13,Z13,A14,B14,C14,D14,E14,F14,G14,H14,I14,J14,K14,L14,M14,N14,O14,P14,Q14,R14,S14,T14,U14,V14,W14,X14,Y14,Z14,A15,B15,C15,D15,E15,F15,G15,H15,I15,J15,K15,L15,M15,N15,O15,P15,Q15,R15,S15,T15,U15,V15,W15,X15,Y15,Z15,A16,B16,C16,D16,E16,F16,G16,H16,I16,J16,K16,L16,M16,N16,O16,P16,Q16,R16,S16,T16,U16,V16,W16,X16,Y16,Z16,A17,B17,C17,D17,E17,F17,G17,H17,I17,J17,K17,L17,M17,N17,O17,P17,Q17,R17,S17,T17,U17,V17,W17,X17,Y17,Z17,A18,B18,C18,D18,E18,F18,G18,H18,I18,J18,K18,L18,M18,N18,O18,P18,Q18,R18,S18,T18,U18,V18,W18,X18,Y18,Z18,A19,B19,C19,D19,E19,F19,G19,H19,I19,J19,K19,L19,M19,N19,O19,P19,Q19,R19,S19,T19,U19,V19,W19,X19,Y19,Z19,A20,B20,C20,D20,E20,F20,G20,H20,I20,J20,K20,L20,M20,N20,O20,P20,Q20,R20,S20,T20,U20,V20,W20,X20,Y20,Z20,A21,B21,C21,D21,E21,F21,G21,H21,I21,J21,K21,L21,M21,N21,O21,P21,Q21,R21,S21,T21,U21,V21,W21,X21,Y21,Z21,A22,B22,C22,D22,E22,F22,G22,H22,I22,J22,K22,L22,M22,N22,O22,P22,Q22,R22,S22,T22,U22,V22,W22,X22,Y22,Z22,A23,B23,C23,D23,E23,F23,G23,H23,I23,J23,K23,L23,M23,N23,O23,P23,Q23,R23,S23,T23,U23,V23,W23,X23,Y23,Z23,A24,B24,C24,D24,E24,F24,G24,H24,I24,J24,K24,L24,M24,N24,O24,P24,Q24,R24,S24,T24,U24,V24,W24,X24,Y24,Z24,A25,B25,C25,D25,E25,F25,G25,H25,I25,J25,K25,L25,M25,N25,O25,P25,Q25,R25,S25,T25,U25,V25,W25,X25,Y25,Z25,A26,B26,C26,D26,E26,F26,G26,H26,I26,J26,K26,L26,M26,N26,O26,P26,Q26,R26,S26,T26,U26,V26,W26,X26,Y26,Z26,A27,B27,C27,D27,E27,F27,G27,H27,I27,J27,K27,L27,M27,N27,O27,P27,Q27,R27,S27,T27,U27,V27,W27,X27,Y27,Z27,A28,B28,C28,D28,E28,F28,G28,H28,I28,J28,K28,L28,M28,N28,O28,P28,Q28,R28,S28,T28,U28,V28,W28,X28,Y28,Z28,A29,B29,C29,D29,E29,F29,G29,H29,I29,J29,K29,L29,M29,N29,O29,P29,Q29,R29,S29,T29,U29,V29,W29,X29,Y29,Z29,A30,B30,C30,D30,E30,F30,G30,H30,I30,J30,K30,L30,M30,N30,O30,P30,Q30,R30,S30,T30,U30,V30,W30,X30,Y30,Z30,A31,B31,C31,D31,E31,F31,G31,H31,I31,J31,K31,L31,M31,N31,O31,P31,Q31,R31,S31,T31,U31,V31,W31,X31,Y31,Z31,A32,B32,C32,D32,E32,F32,G32,H32,I32,J32,K32,L32,M32,N32,O32,P32,Q32,R32,S32,T32,U32,V32,W32,X32,Y32,Z32,A33,B33,C33,D33,E33,F33,G33,H33,I33,J33,K33,L33,M33,N33,O33,P33,Q33,R33,S33,T33,U33,V33,W33,X33,Y33,Z33,A34,B34,C34,D34,E34,F34,G34,H34,I34,J34,K34,L34,M34,N34,O34,P34,Q34,R34,S34,T34,U34,V34,W34,X34,Y34,Z34,A35,B35,C35,D35,E35,F35,G35,H35,I35,J35,K35,L35,M35,N35,O35,P35,Q35,R35,S35,T35,U35,V35,W35,X35,Y35,Z35,A36,B36,C36,D36,E36,F36,G36,H36,I36,J36,K36,L36,M36,N36,O36,P36,Q36,R36,S36,T36,U36,V36,W36,X36,Y36,Z36,A37,B37,C37,D37,E37,F37,G37,H37,I37,J37,K37,L37,M37,N37,O37,P37,Q37,R37,S37,T37,U37,V37,W37,X37,Y37,Z37,A38,B38,C38,D38,E38,F38,G38,H38,I38,J38,K38,L38,M38,N38,O38,P38,Q38,R38,S38,T38,U38,V38,W38,X38,Y38,Z38,A39,B39,C39,D39,E39,F39,G39,H39,I39,J39,K39,L39],M39).
r10(A,B) <=> r9(A,B).
r9([T19,S19,R19,Q19,P19,O19,N19,M19,L19,K19,J19,I19,H19,G19,F19,E19,D19,C19,B19,A19,Z18,Y18,X18,W18,V18,U18,T18,S18,R18,Q18,P18,O18,N18,M18,L18,K18,J18,I18,H18,G18,F18,E18,D18,C18,B18,A18,Z17,Y17,X17,W17,V17,U17,T17,S17,R17,Q17,P17,O17,N17,M17,L17,K17,J17,I17,H17,G17,F17,E17,D17,C17,B17,A17,Z16,Y16,X16,W16,V16,U16,T16,S16,R16,Q16,P16,O16,N16,M16,L16,K16,J16,I16,H16,G16,F16,E16,D16,C16,B16,A16,Z15,Y15,X15,W15,V15,U15,T15,S15,R15,Q15,P15,O15,N15,M15,L15,K15,J15,I15,H15,G15,F15,E15,D15,C15,B15,A15,Z14,Y14,X14,W14,V14,U14,T14,S14,R14,Q14,P14,O14,N14,M14,L14,K14,J14,I14,H14,G14,F14,E14,D14,C14,B14,A14,Z13,Y13,X13,W13,V13,U13,T13,S13,R13,Q13,P13,O13,N13,M13,L13,K13,J13,I13,H13,G13,F13,E13,D13,C13,B13,A13,Z12,Y12,X12,W12,V12,U12,T12,S12,R12,Q12,P12,O12,N12,M12,L12,K12,J12,I12,H12,G12,F12,E12,D12,C12,B12,A12,Z11,Y11,X11,W11,V11,U11,T11,S11,R11,Q11,P11,O11,N11,M11,L11,K11,J11,I11,H11,G11,F11,E11,D11,C11,B11,A11,Z10,Y10,X10,W10,V10,U10,T10,S10,R10,Q10,P10,O10,N10,M10,L10,K10,J10,I10,H10,G10,F10,E10,D10,C10,B10,A10,Z9,Y9,X9,W9,V9,U9,T9,S9,R9,Q9,P9,O9,N9,M9,L9,K9,J9,I9,H9,G9,F9,E9,D9,C9,B9,A9,Z8,Y8,X8,W8,V8,U8,T8,S8,R8,Q8,P8,O8,N8,M8,L8,K8,J8,I8,H8,G8,F8,E8,D8,C8,B8,A8,Z7,Y7,X7,W7,V7,U7,T7,S7,R7,Q7,P7,O7,N7,M7,L7,K7,J7,I7,H7,G7,F7,E7,D7,C7,B7,A7,Z6,Y6,X6,W6,V6,U6,T6,S6,R6,Q6,P6,O6,N6,M6,L6,K6,J6,I6,H6,G6,F6,E6,D6,C6,B6,A6,Z5,Y5,X5,W5,V5,U5,T5,S5,R5,Q5,P5,O5,N5,M5,L5,K5,J5,I5,H5,G5,F5,E5,D5,C5,B5,A5,Z4,Y4,X4,W4,V4,U4,T4,S4,R4,Q4,P4,O4,N4,M4,L4,K4,J4,I4,H4,G4,F4,E4,D4,C4,B4,A4,Z3,Y3,X3,W3,V3,U3,T3,S3,R3,Q3,P3,O3,N3,M3,L3,K3,J3,I3,H3,G3,F3,E3,D3,C3,B3,A3,Z2,Y2,X2,W2,V2,U2,T2,S2,R2,Q2,P2,O2,N2,M2,L2,K2,J2,I2,H2,G2,F2,E2,D2,C2,B2,A2,Z1,Y1,X1,W1,V1,U1,T1,S1,R1,Q1,P1,O1,N1,M1,L1,K1,J1,I1,H1,G1,F1,E1,D1,C1,B1,A1,Z,Y,X,W,V,U,T,S,R,Q,P,O,N,M,L,K,J,I,H,G,F,E,D,C|A],U19) <=> r8(A,B),
        append(B,[C,D,E,F,G,H,I,J,K,L,M,N,O,P,Q,R,S,T,U,V,W,X,Y,Z,A1,B1,C1,D1,E1,F1,G1,H1,I1,J1,K1,L1,M1,N1,O1,P1,Q1,R1,S1,T1,U1,V1,W1,X1,Y1,Z1,A2,B2,C2,D2,E2,F2,G2,H2,I2,J2,K2,L2,M2,N2,O2,P2,Q2,R2,S2,T2,U2,V2,W2,X2,Y2,Z2,A3,B3,C3,D3,E3,F3,G3,H3,I3,J3,K3,L3,M3,N3,O3,P3,Q3,R3,S3,T3,U3,V3,W3,X3,Y3,Z3,A4,B4,C4,D4,E4,F4,G4,H4,I4,J4,K4,L4,M4,N4,O4,P4,Q4,R4,S4,T4,U4,V4,W4,X4,Y4,Z4,A5,B5,C5,D5,E5,F5,G5,H5,I5,J5,K5,L5,M5,N5,O5,P5,Q5,R5,S5,T5,U5,V5,W5,X5,Y5,Z5,A6,B6,C6,D6,E6,F6,G6,H6,I6,J6,K6,L6,M6,N6,O6,P6,Q6,R6,S6,T6,U6,V6,W6,X6,Y6,Z6,A7,B7,C7,D7,E7,F7,G7,H7,I7,J7,K7,L7,M7,N7,O7,P7,Q7,R7,S7,T7,U7,V7,W7,X7,Y7,Z7,A8,B8,C8,D8,E8,F8,G8,H8,I8,J8,K8,L8,M8,N8,O8,P8,Q8,R8,S8,T8,U8,V8,W8,X8,Y8,Z8,A9,B9,C9,D9,E9,F9,G9,H9,I9,J9,K9,L9,M9,N9,O9,P9,Q9,R9,S9,T9,U9,V9,W9,X9,Y9,Z9,A10,B10,C10,D10,E10,F10,G10,H10,I10,J10,K10,L10,M10,N10,O10,P10,Q10,R10,S10,T10,U10,V10,W10,X10,Y10,Z10,A11,B11,C11,D11,E11,F11,G11,H11,I11,J11,K11,L11,M11,N11,O11,P11,Q11,R11,S11,T11,U11,V11,W11,X11,Y11,Z11,A12,B12,C12,D12,E12,F12,G12,H12,I12,J12,K12,L12,M12,N12,O12,P12,Q12,R12,S12,T12,U12,V12,W12,X12,Y12,Z12,A13,B13,C13,D13,E13,F13,G13,H13,I13,J13,K13,L13,M13,N13,O13,P13,Q13,R13,S13,T13,U13,V13,W13,X13,Y13,Z13,A14,B14,C14,D14,E14,F14,G14,H14,I14,J14,K14,L14,M14,N14,O14,P14,Q14,R14,S14,T14,U14,V14,W14,X14,Y14,Z14,A15,B15,C15,D15,E15,F15,G15,H15,I15,J15,K15,L15,M15,N15,O15,P15,Q15,R15,S15,T15,U15,V15,W15,X15,Y15,Z15,A16,B16,C16,D16,E16,F16,G16,H16,I16,J16,K16,L16,M16,N16,O16,P16,Q16,R16,S16,T16,U16,V16,W16,X16,Y16,Z16,A17,B17,C17,D17,E17,F17,G17,H17,I17,J17,K17,L17,M17,N17,O17,P17,Q17,R17,S17,T17,U17,V17,W17,X17,Y17,Z17,A18,B18,C18,D18,E18,F18,G18,H18,I18,J18,K18,L18,M18,N18,O18,P18,Q18,R18,S18,T18,U18,V18,W18,X18,Y18,Z18,A19,B19,C19,D19,E19,F19,G19,H19,I19,J19,K19,L19,M19,N19,O19,P19,Q19,R19,S19,T19],U19).
r9(A,B) <=> r8(A,B).
r8([X9,W9,V9,U9,T9,S9,R9,Q9,P9,O9,N9,M9,L9,K9,J9,I9,H9,G9,F9,E9,D9,C9,B9,A9,Z8,Y8,X8,W8,V8,U8,T8,S8,R8,Q8,P8,O8,N8,M8,L8,K8,J8,I8,H8,G8,F8,E8,D8,C8,B8,A8,Z7,Y7,X7,W7,V7,U7,T7,S7,R7,Q7,P7,O7,N7,M7,L7,K7,J7,I7,H7,G7,F7,E7,D7,C7,B7,A7,Z6,Y6,X6,W6,V6,U6,T6,S6,R6,Q6,P6,O6,N6,M6,L6,K6,J6,I6,H6,G6,F6,E6,D6,C6,B6,A6,Z5,Y5,X5,W5,V5,U5,T5,S5,R5,Q5,P5,O5,N5,M5,L5,K5,J5,I5,H5,G5,F5,E5,D5,C5,B5,A5,Z4,Y4,X4,W4,V4,U4,T4,S4,R4,Q4,P4,O4,N4,M4,L4,K4,J4,I4,H4,G4,F4,E4,D4,C4,B4,A4,Z3,Y3,X3,W3,V3,U3,T3,S3,R3,Q3,P3,O3,N3,M3,L3,K3,J3,I3,H3,G3,F3,E3,D3,C3,B3,A3,Z2,Y2,X2,W2,V2,U2,T2,S2,R2,Q2,P2,O2,N2,M2,L2,K2,J2,I2,H2,G2,F2,E2,D2,C2,B2,A2,Z1,Y1,X1,W1,V1,U1,T1,S1,R1,Q1,P1,O1,N1,M1,L1,K1,J1,I1,H1,G1,F1,E1,D1,C1,B1,A1,Z,Y,X,W,V,U,T,S,R,Q,P,O,N,M,L,K,J,I,H,G,F,E,D,C|A],Y9) <=> r7(A,B),
        append(B,[C,D,E,F,G,H,I,J,K,L,M,N,O,P,Q,R,S,T,U,V,W,X,Y,Z,A1,B1,C1,D1,E1,F1,G1,H1,I1,J1,K1,L1,M1,N1,O1,P1,Q1,R1,S1,T1,U1,V1,W1,X1,Y1,Z1,A2,B2,C2,D2,E2,F2,G2,H2,I2,J2,K2,L2,M2,N2,O2,P2,Q2,R2,S2,T2,U2,V2,W2,X2,Y2,Z2,A3,B3,C3,D3,E3,F3,G3,H3,I3,J3,K3,L3,M3,N3,O3,P3,Q3,R3,S3,T3,U3,V3,W3,X3,Y3,Z3,A4,B4,C4,D4,E4,F4,G4,H4,I4,J4,K4,L4,M4,N4,O4,P4,Q4,R4,S4,T4,U4,V4,W4,X4,Y4,Z4,A5,B5,C5,D5,E5,F5,G5,H5,I5,J5,K5,L5,M5,N5,O5,P5,Q5,R5,S5,T5,U5,V5,W5,X5,Y5,Z5,A6,B6,C6,D6,E6,F6,G6,H6,I6,J6,K6,L6,M6,N6,O6,P6,Q6,R6,S6,T6,U6,V6,W6,X6,Y6,Z6,A7,B7,C7,D7,E7,F7,G7,H7,I7,J7,K7,L7,M7,N7,O7,P7,Q7,R7,S7,T7,U7,V7,W7,X7,Y7,Z7,A8,B8,C8,D8,E8,F8,G8,H8,I8,J8,K8,L8,M8,N8,O8,P8,Q8,R8,S8,T8,U8,V8,W8,X8,Y8,Z8,A9,B9,C9,D9,E9,F9,G9,H9,I9,J9,K9,L9,M9,N9,O9,P9,Q9,R9,S9,T9,U9,V9,W9,X9],Y9).
r8(A,B) <=> r7(A,B).
r7([Z4,Y4,X4,W4,V4,U4,T4,S4,R4,Q4,P4,O4,N4,M4,L4,K4,J4,I4,H4,G4,F4,E4,D4,C4,B4,A4,Z3,Y3,X3,W3,V3,U3,T3,S3,R3,Q3,P3,O3,N3,M3,L3,K3,J3,I3,H3,G3,F3,E3,D3,C3,B3,A3,Z2,Y2,X2,W2,V2,U2,T2,S2,R2,Q2,P2,O2,N2,M2,L2,K2,J2,I2,H2,G2,F2,E2,D2,C2,B2,A2,Z1,Y1,X1,W1,V1,U1,T1,S1,R1,Q1,P1,O1,N1,M1,L1,K1,J1,I1,H1,G1,F1,E1,D1,C1,B1,A1,Z,Y,X,W,V,U,T,S,R,Q,P,O,N,M,L,K,J,I,H,G,F,E,D,C|A],A5) <=> r6(A,B),
        append(B,[C,D,E,F,G,H,I,J,K,L,M,N,O,P,Q,R,S,T,U,V,W,X,Y,Z,A1,B1,C1,D1,E1,F1,G1,H1,I1,J1,K1,L1,M1,N1,O1,P1,Q1,R1,S1,T1,U1,V1,W1,X1,Y1,Z1,A2,B2,C2,D2,E2,F2,G2,H2,I2,J2,K2,L2,M2,N2,O2,P2,Q2,R2,S2,T2,U2,V2,W2,X2,Y2,Z2,A3,B3,C3,D3,E3,F3,G3,H3,I3,J3,K3,L3,M3,N3,O3,P3,Q3,R3,S3,T3,U3,V3,W3,X3,Y3,Z3,A4,B4,C4,D4,E4,F4,G4,H4,I4,J4,K4,L4,M4,N4,O4,P4,Q4,R4,S4,T4,U4,V4,W4,X4,Y4,Z4],A5).
r7(A,B) <=> r6(A,B).
r6([N2,M2,L2,K2,J2,I2,H2,G2,F2,E2,D2,C2,B2,A2,Z1,Y1,X1,W1,V1,U1,T1,S1,R1,Q1,P1,O1,N1,M1,L1,K1,J1,I1,H1,G1,F1,E1,D1,C1,B1,A1,Z,Y,X,W,V,U,T,S,R,Q,P,O,N,M,L,K,J,I,H,G,F,E,D,C|A],O2) <=> r5(A,B),
        append(B,[C,D,E,F,G,H,I,J,K,L,M,N,O,P,Q,R,S,T,U,V,W,X,Y,Z,A1,B1,C1,D1,E1,F1,G1,H1,I1,J1,K1,L1,M1,N1,O1,P1,Q1,R1,S1,T1,U1,V1,W1,X1,Y1,Z1,A2,B2,C2,D2,E2,F2,G2,H2,I2,J2,K2,L2,M2,N2],O2).
r6(A,B) <=> r5(A,B).
r5([H1,G1,F1,E1,D1,C1,B1,A1,Z,Y,X,W,V,U,T,S,R,Q,P,O,N,M,L,K,J,I,H,G,F,E,D,C|A],I1) <=> r4(A,B),
        append(B,[C,D,E,F,G,H,I,J,K,L,M,N,O,P,Q,R,S,T,U,V,W,X,Y,Z,A1,B1,C1,D1,E1,F1,G1,H1],I1).
r5(A,B) <=> r4(A,B).
r4([R,Q,P,O,N,M,L,K,J,I,H,G,F,E,D,C|A],S) <=> r3(A,B),
        append(B,[C,D,E,F,G,H,I,J,K,L,M,N,O,P,Q,R],S).
r4(A,B) <=> r3(A,B).
r3([J,I,H,G,F,E,D,C|A],K) <=> r2(A,B),append(B,[C,D,E,F,G,H,I,J],K).
r3(A,B) <=> r2(A,B).
r2([F,E,D,C|A],G) <=> r1(A,B),append(B,[C,D,E,F],G).
r2(A,B) <=> r1(A,B).
r1([D,C|A],E) <=> r0(A,B),append(B,[C,D],E).
r1(A,B) <=> r0(A,B).
r0([C|A],D) <=> r(A,B),append(B,[C],D).
r0(A,B) <=> r(A,B).
r([],D) <=> D=[].
\end{verbatim}
} %small

\label{lastpage}

\end{document}